\documentclass[floatfix,%
 reprint,
 amsmath,amssymb,
 aps,
]{revtex4-2}
\usepackage{graphicx}
\usepackage{dcolumn}
\usepackage{bm}
\usepackage{color}
\usepackage{xcolor}
\usepackage{natbib}
\usepackage{comment}
\DeclareGraphicsExtensions{.pdf,.eps,.png,.jpg,.mps}
\usepackage{float}

\begin{document} 

\title{Perfect continuous-variable quantum microcombs}
\author
{Kangkang Li$^{1,\ast,\dagger}$, Yue Wang$^{1,\ast}$, Ze Wang$^{1,\ast}$, Xin Zhou$^{2,\ast}$, Jincheng Li$^{2,3}$, Yinke Cheng$^{1}$, Binyan Wu$^{1}$, Qihuang Gong$^{1,4,5,6}$, Bei-Bei Li$^{2,\dagger}$, Qi-Fan Yang$^{1,4,5,6,\dagger}$
\\$^{1}$State Key Laboratory for Artificial Microstructure and Mesoscopic Physics and Frontiers Science Center for Nano-optoelectronics School of Physics, Peking University, Beijing, 100871, China\\
$^2$Beijing National Laboratory for Condensed Matter Physics, Institute of Physics, Chinese Academy of Sciences, Beijing, 100190, China\\
$^3$School of Space and Earth Sciences, Beihang University, Beijing 100191, China\\
$^4$Collaborative Innovation Center of Extreme Optics, Shanxi University, Taiyuan, 030006, China\\
$^5$Peking University Yangtze Delta Institute of Optoelectronics, Nantong, Jiangsu, 226010, China\\
$^6$Hefei National Laboratory, Hefei, 230088, China\\
$^{\ast}$These authors contributed equally to this work.\\
$^{\dagger}$Corresponding author: kangkangli@pku.edu.cn; libeibei@iphy.ac.cn; leonardoyoung@pku.edu.cn.
}

\begin{abstract}
Quantum microcombs generated in high-$Q$ microresonators provide compact, multiplexed sources of entangled modes for continuous-variable (CV) quantum information processing. While deterministic generation of CV states via Kerr-induced two-mode squeezing has been demonstrated, achieving spectrally uniform squeezing remains challenging because of asymmetry and anomalies in the dispersion profile. Here we overcome these limitations by combining a microresonator with an engineered mode spectrum and optimized pump conditions. We realize a CV quantum microcomb comprising 14 independent two-mode squeezed states, each exhibiting more than 4~dB of raw squeezing (up to 4.3~dB) across a 0.7~THz bandwidth. This uniform, high-performance quantum resource represents a key step toward scalable, integrated CV quantum technologies operating beyond classical limits.
\end{abstract}

\maketitle

\emph{Introduction.—}
Continuous-variable (CV) quantum optics exploits the quadrature amplitudes of light to encode and manipulate quantum states \cite{braunstein2005quantum, menicucci2006universal}. Over the past two decades, the CV framework has evolved from early demonstrations of squeezed-light interferometry \cite{grangier1987squeezed,xiao1987precision,schnabel2017squeezed} to universal, measurement-based quantum computation on large cluster states \cite{menicucci2006universal,van2007building,menicucci2008one,menicucci2014fault,fukui2018high}. Quadrature squeezing and entanglement have enabled enhanced sensitivity in gravitational-wave detectors \cite{aasi2013enhanced,tse2019quantum}, optomechanical force sensors \cite{li2018quantum}, and quantum-enhanced spectroscopy \cite{polzik1992spectroscopy,herman2025squeezed}, and underpin encodings suitable for quantum error correction and fault-tolerant processing \cite{konno2024logical,larsen2025integrated}. A central challenge for practical deployment is to miniaturize these capabilities onto photonic chips while preserving strong squeezing and supporting large-scale multiplexing in frequency \cite{pysher2011parallel,chen2014experimental,roslund2014wavelength,medeiros2014full,cai2017multimode,roh2025generation}, time \cite{yokoyama2013ultra,asavanant2019generation,larsen2019deterministic,madsen2022quantum}, or path \cite{su2013gate,arrazola2021quantum}.

Quantum microcombs—optical frequency combs generated via parametric nonlinearities in high-$Q$ microresonators—are a promising platform to meet these requirements \cite{kues2019quantum}. Discrete-variable microcombs employ spontaneous four-wave mixing to generate multiplexed photon pairs \cite{reimer2016generation,kues2017chip,imany2018,steiner2021ultrabright,guidry2022quantum,fan2023multi}. In contrast, CV quantum microcombs utilize deterministic two-mode squeezing to produce frequency-multiplexed Einstein-Podolsky-Rosen (EPR) pairs \cite{vaidya2020broadband,yang2021squeezed,Jahanbozorgi2022,jahanbozorgi2023generation,liu2025wafer}, which can be programmed into large cluster-state graphs \cite{wang2025large,jia2025continuous}. The performance of such CV microcomb systems is governed by the squeezing level (SL), defined as the quadrature-noise reduction below the shot-noise level, and by the spectral uniformity of that SL across many mode pairs. State-of-the-art on-chip CV microcombs have demonstrated two-mode squeezing up to 5.6~dB in bright (above-threshold) configurations \cite{dutt2015chip,dutt2016tunable,shen2025strong,shen2025highly} and up to 3~dB in vacuum (below-threshold) states \cite{wang2025large,liu2025wafer}. 
The SLs typically exhibit substantial spectral variation, which limits the number of usable EPR pairs \cite{yang2021squeezed,Jahanbozorgi2022,jahanbozorgi2023generation,wang2025large}.

\begin{figure}[b]
\centering
\includegraphics[width=1.0\linewidth]{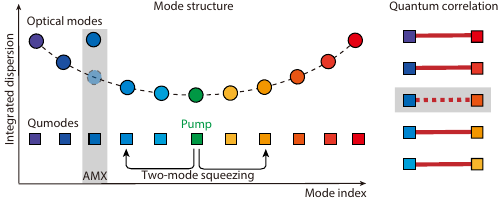}
\caption{\textbf{Mode structure and quantum correlations in a CV quantum microcomb.} Schematic integrated dispersion of a microresonator, showing the distribution of optical modes and the corresponding frequency-multiplexed qumodes used for two-mode squeezing. Gray shaded regions indicate modes strongly perturbed by avoided mode crossings (AMXs), where dispersion asymmetry degrades quantum correlations.}
\label{figure1}
\end{figure}

\begin{figure}
\centering
\includegraphics[width=1.0\linewidth]{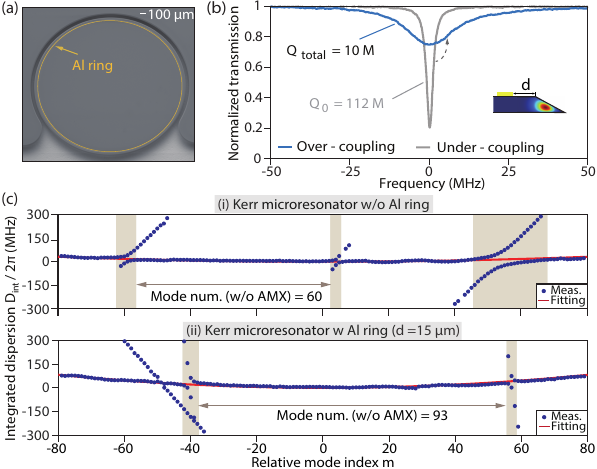}
\caption{\textbf{Spectrally purified microresonator.} (a) Scanning electron micrograph of a silica microdisk with a 10~$\mu$m-wide Al ring deposited near the disk edge. (b) Transmission spectrum of the fundamental transverse mode family. Inset: simulated mode profile in the disk cross section. (c) Integrated dispersion of the fundamental family without (i) and with (ii) the Al ring. The metal ring selectively damps higher-order families and yields an AMX-free window around 1543.2~nm.}
\label{figure2}
\end{figure}

In this work, we demonstrate a spectrally uniform, vacuum-state CV quantum microcomb that overcomes these limitations. Our approach is based on two key ingredients: (i) engineering the microresonator to realize an almost ideal, single-family mode structure in which higher-order transverse modes and their associated AMXs are strongly suppressed, and (ii) optimizing the pump condition to maximize the bandwidth over which near-maximal squeezing is achieved. With this strategy, we generate a CV microcomb that supports more than 14 EPR pairs, each exhibiting over 4~dB of raw squeezing (up to 4.3~dB) across a 0.7~THz span.

\emph{Theory.—}
We first formalize the concept of a CV quantum microcomb and analyze how dispersion asymmetry impacts quantum correlations [Fig. 1]. For simplicity, the frequency units are normalized to half the optical damping rate (assumed mode-independent). Consider a whispering-gallery microresonator with transverse mode frequencies $\omega_k$, where the integer index $k$ is referenced to the pumped resonance ($k=0$). The integrated dispersion is defined as $D_{\mathrm{int}}(k) = \omega_k - \omega_0 - k D_1$, where $D_1$ is the free-spectral range at the pump frequency. 
The pump laser is red detuned from the cold-cavity resonance by $\zeta_0>0$. 
Quantum-correlated pairs of frequency modes (qumodes) are generated symmetrically at $\pm k$ via degenerate four-wave mixing. Let $\zeta_k$ denote the detuning of the qumode centered near mode $k$ from the corresponding optical resonance. The phase-matching condition for the EPR pair $(\hat{a}_k,\hat{a}_{-k})$ can be written as $\zeta_k + \zeta_{-k} - D_{\mathrm{int}}(k) - D_{\mathrm{int}}(-k) = 2\zeta_0$.

The linearized quantum-Langevin dynamics for the creation and annihilation operators of the qumodes are
\begin{equation}
\begin{split}
\frac{d\hat a_k}{d\tau}&=[-1-i(\overline\zeta_k+\Delta_k-2\alpha)]\hat a_k+i\alpha\hat a^{\dagger}_{-k}+\hat W_k\\
\frac{d\hat a^{\dagger}_{-k}}{d\tau}&=[-1+i(\overline\zeta_k-\Delta_k-2\alpha)]\hat a^{\dagger}_{-k}-i\alpha\hat a_k+\hat W^{\dagger}_{-k}
\end{split}\label{Eq:CM}
\end{equation}
where $\tau$ is the time normalized to the cavity photon lifetime, $\alpha$ is the dimensionless parametric coupling strength proportional to the intracavity power, and the Langevin forces $\hat{W}_{\pm k}$ describe vacuum input noise. To remain below the oscillation threshold, we require $\alpha<1$. We introduce the average and asymmetric detunings
\begin{equation}
\begin{aligned}
\overline{\zeta}_k &= \frac{\zeta_k+\zeta_{-k}}{2}
= \frac{D_{\mathrm{int}}(k)+D_{\mathrm{int}}(-k)}{2} + \zeta_0,
\\
\Delta_k &= \frac{\zeta_k-\zeta_{-k}}{2}.
\end{aligned}
\end{equation}

\begin{figure}[t]
\centering
\includegraphics{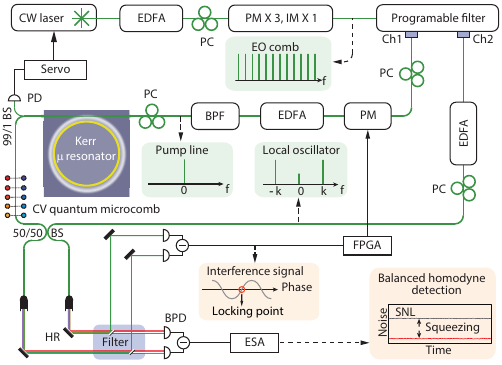}
\caption{\textbf{Experimental setup.} CW laser: continuous-wave laser; EO comb: electro-optic comb; EDFA: erbium-doped fiber amplifier; PM/IM: phase/intensity modulators; BPF: band-pass filter; PC: polarization controller; BS: beam splitter; HR: highly reflective mirror; PD: photodetector; BPD: balanced photodetector; ESA: electrical spectrum analyzer; FPGA: field-programmable gate array.}
\label{figure3}
\end{figure}

The detected field operators $\hat{c}_k$ are related to the intracavity fields by the standard input-output relations. We define the amplitude and phase quadratures of the detected modes as $\hat{x}_k = \hat{c}_k + \hat{c}_k^{\dagger}$, $\hat{p}_k = -i(\hat{c}_k - \hat{c}_k^{\dagger})$.
Including a total detection efficiency $\eta$, the shot-noise–normalized variance of the EPR sum quadrature $\hat{x}_k + \hat{x}_{-k}$ is found to be
\begin{equation}
\mathrm{Var}(\hat{x}_k+\hat{x}_{-k})
= 1 - \eta\,\frac{2}{1+\sqrt{1+\Lambda_k/(4\alpha^2)}},
\label{eq:Var}
\end{equation}
with 
\begin{equation}
    \Lambda_k = \bigl[1-\alpha^2+(\overline{\zeta}_k-2\alpha)^2-\Delta_k^2\bigr]^2 + 4\Delta_k^2.
\end{equation}
The squeezing level for the pair $(k,-k)$ is then given by $10\log_{10}\bigl[\mathrm{Var}(\hat{x}_k+\hat{x}_{-k})\bigr]$ in dB.
For small red detuning satisfying $(\overline{\zeta}_k-2\alpha)^2<1+\alpha^2$, the minimum of $\Lambda_k$ occurs at symmetric detuning $\Delta_k=0$. This condition can be fulfilled in microresonators whose dispersion is dominated by second-order terms, $D_{\mathrm{int}}(k)\simeq D_2 k^2/2$, such that the integrated dispersion is nearly parabolic and the qumodes form an equidistant frequency comb. In practice, however, odd-order dispersion and AMXs arising from coupling to higher-order transverse families break the symmetry and generate nonzero $\Delta_k$ in otherwise equidistant quantum microcombs. In particular, AMXs can shift resonances by many linewidths, creating regimes with $\Delta_k\gg1$ that drastically suppress quantum correlations for the corresponding EPR pairs.

\begin{figure*}
\centering
\includegraphics[width=1\linewidth]{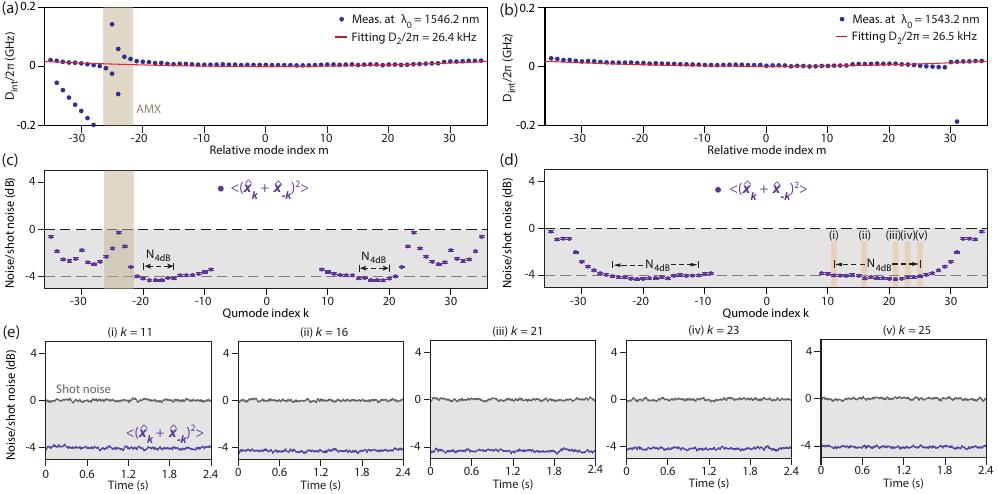}
\caption{\textbf{CV microcomb with and without AMX.} (a,b) Measured integrated dispersion near 1546.2~nm (AMX near $k=-24$) and 1543.2~nm (AMX-free). (c,d) Shot-noise–normalized two-mode quadrature variances for EPR pairs at the two pump wavelengths. (e) Representative homodyne traces at 0.5~MHz (resolution bandwidth 100~kHz, video bandwidth 10~Hz, 2.4~s sweep) for $(11,-11)$, $(16,-16)$, $(21,-21)$, $(23,-23)$, and $(25,-25)$. Electronic noise of the photodetector is subtracted; the shot-noise level is shown in gray.}
\label{figure4}
\end{figure*}

\emph{Perfect mode structure.—} High-$Q$ microresonators often support many transverse mode families. Large cross sections that reduce sidewall scattering and increase $Q$ simultaneously facilitate the existence of higher-order modes. In silica wedge disks, tens of high-$Q$ families are commonly observed \cite{lee2012chemically}. Coupling between these families produces frequent AMXs, which break the spectral symmetry required for uniform squeezing.

To suppress AMXs, we engineer the transverse-mode structure so that higher-order families are strongly damped and do not appreciably couple to the fundamental family (see Supplementary Material). The device in Fig. 2(a) is a silica disk with a free-spectral range of 25~GHz. A 10~$\mu$m-wide Al ring is deposited on the upper surface, positioned 15~$\mu$m inward from the wedge. Owing to its overlap with higher-order modes, the metal ring introduces selective loss that reduces their $Q$ factors, while the fundamental family retains $Q_i>1.1\times10^8$ and a loaded $Q\simeq1.0\times10^7$, corresponding to an external coupling efficiency near 90\% [Fig. 2(b)]. 

The measured integrated dispersion of the fundamental family is nearly parabolic. Two weak residual families generate localized AMXs, but between these perturbations we obtain an AMX-free window of 93 consecutive modes (approximately 2~THz) centered at 1543.2~nm [Fig. 2(c)]. By contrast, devices fabricated without the selective-loss treatment exhibit substantially more AMXs, owing to the persistence of many high-$Q$ higher-order families.

The experimental setup is shown in Fig. 3. A programmable electro-optic (EO) frequency comb is generated by phase and amplitude modulation of a continuous-wave laser \cite{ishizawa2011generation,metcalf2013high}. One comb line is spectrally filtered and used to pump the microresonator, while selected sidebands, together with a small fraction of the pump, serve as local oscillators (LOs) for balanced homodyne detection. The pump is coupled into the resonator via a tapered fiber, and a portion of the transmitted pump is tapped to generate an error signal for active frequency stabilization. The CV microcomb and LO fields interfere on a 50:50 coupler, and the pump line is rejected by a narrowband filter before detection. The path-length difference between the comb and LO arms is stabilized using a phase modulator; an FPGA implements wide-dynamic-range feedback based on the interference of the filtered pump. Quadrature-noise spectra are acquired with an electrical spectrum analyzer. The total detection efficiency, including coupling, filtering, and photodiode quantum efficiency, is $\eta\simeq0.70$.

To quantify the impact of AMXs, we compare pumping at 1546.2~nm and 1543.2~nm. The measured $D_{\mathrm{int}}(k)$ is well described by a quadratic with $D_2/2\pi\approx26.5$~kHz. When the pump is tuned near 1546.2~nm, an AMX appears around $k=-24$ [Fig. 4(a)], while at 1543.2~nm the fundamental family remains free of discernible AMXs over 71 consecutive modes [Fig. 4(b)]. We record the shot-noise–normalized variances of $\hat{x}_k+\hat{x}_{-k}$ across $k=\pm9$ to $\pm35$. For the 1546.2~nm pump, the maximum SL approaches 4.26~dB but collapses in the vicinity of the AMX [Fig. 4(c)]; at the mode with the largest deviation from the quadratic dispersion ($k=24$), squeezing nearly vanishes, consistent with Eq. (3). In contrast, for the 1543.2~nm pump, the SLs are nearly uniform across the measured band [Fig. 4(d)]. Time-series traces of the shot-noise–normalized variance for selected pairs are shown in Fig. 4(e). Representative homodyne spectra at $k=\pm21$ exhibit $4.32\pm0.08$~dB of squeezing in $\hat{x}_{21}+\hat{x}_{-21}$. In total, 14 EPR pairs show $\mathrm{SL}>4$~dB.

\emph{Perfect pumping condition.—}
Finally, we determine the pump conditions that maximize the bandwidth of the squeezed CV quantum microcomb, quantified by the number of EPR pairs with nearly uniform SL. In the absence of AMXs (so that $\Delta_k\simeq0$), Equation (3) predicts that maximal squeezing for a given pair occurs near $\zeta_k=2\alpha$. If $\zeta_k$ remains within the interval $\alpha\le\zeta_k\le3\alpha$, the quadrature-noise reduction $1-\mathrm{Var}(\hat{x}_k+\hat{x}_{-k})$ stays above 92\% of its maximum value, corresponding to a degradation of less than 0.65~dB from the optimal SL at a total detection efficiency $\eta\leq 0.7$. We therefore define this interval as the \emph{uniform squeezing regime} of the CV quantum microcomb. For a given device, this regime can be inferred directly from the dispersion profile.

\begin{figure}[h]
\centering
\includegraphics[width=1.0\linewidth]{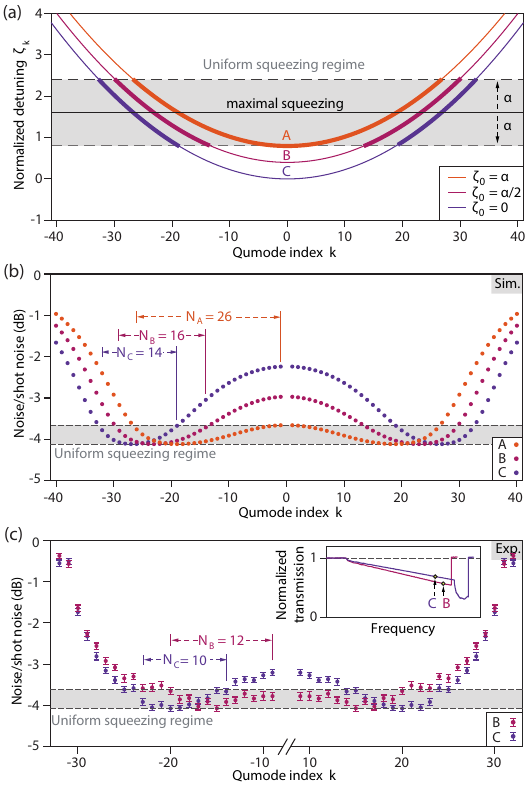}
\caption{\textbf{Parametric control of the microcomb bandwidth.} 
(a) Normalized detunings $\zeta_k$ of different qumodes for three pump detunings: $\zeta_0=\alpha$ (A), $\zeta_0=\alpha/2$ (B), and $\zeta_0=0$ (C). 
(b) Simulated shot-noise–normalized quadrature-noise variances at $\alpha=0.8$ for $\zeta_0=0.8$ (A), $0.4$ (B), and $0$ (C).  
(c) Measured quadrature-noise variances for the same operating points as in (b). 
Inset: transmission spectra at pump powers of 29~$\mu$W (B) and 39~$\mu$W (C); the operating points are indicated by circles.}
\label{figure5}
\end{figure}

Retaining only second-order dispersion, the mode detunings can be approximated as
$\zeta_k \simeq \zeta_0 + \frac{D_2}{2}k^2$. Overlaying this parabola with the uniform squeezing regime gives the range of admissible mode indices. The maximum bandwidth is obtained at the optimal normalized detuning $\zeta_0=\alpha$ [Fig. 5(a)]. In this case, the number of available EPR pairs scales as
\begin{equation}
N_{\mathrm{pairs}}\sim 2\sqrt{\alpha/D_2}.
\end{equation}
Detunings larger than $\zeta_0\simeq1$ are not supported in our system. As $\zeta_0$ is reduced below this optimum, the accessible range of $|k|$ shrinks and the bandwidth decreases.

These trends are confirmed numerically by simulating the SL as a function of pump detuning while adjusting the pump power to keep $\alpha$ fixed. For $\alpha=0.8$, we obtain CV quantum microcombs at different $\zeta_0$ that share nearly the same maximal SL, while the spectral span over which the SL remains nearly uniform increases with $\zeta_0$ [Fig. 5(b)]. In particular, the number of EPR pairs within the uniform squeezing regime increases from 14 to 26 as $\zeta_0$ is raised over the accessible range. 

Experimentally, we verify these predictions at $\alpha=0.8$ for two representative detunings, $\zeta_0=0$ and $\zeta_0=\alpha/2$ [Fig. 5(c)]. This is accomplished by varying the pump power and adjusting the laser detuning while keeping the intracavity power (and hence $\alpha$) fixed (see supplementary materials). The largest detunings predicted to be optimal are experimentally challenging to access because of the thermal stability of the microresonator. Within the accessible range, we find that 10 EPR pairs fall within the uniform squeezing regime for $\zeta_0=0$, increasing to 12 pairs for $\zeta_0=\alpha/2$. This behavior follows the theoretical trend that larger detuning leads to a more uniform distribution of SLs over a broader set of EPR pairs. Consistent with the simulations, the uniform squeezing window also shifts closer to the pump frequency at larger detuning, easing the bandwidth requirements of the LO.

\begin{figure}[h]
\centering
\includegraphics[width=1.0\linewidth]{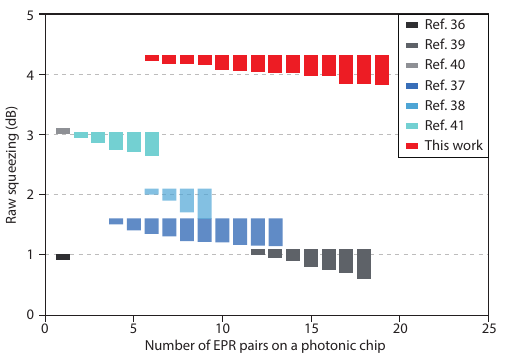}
\caption{\textbf{Comparison of raw squeezing and number of EPR pairs on photonic chips.} 
Vacuum two-mode squeezing realized in $\mathrm{Si}_{3}\mathrm{N}_4$ \cite{vaidya2020broadband,jahanbozorgi2023generation,liu2025wafer} and silica microresonators \cite{yang2021squeezed,Jahanbozorgi2022,wang2025large} are shown for comparison. Bars indicate the range of SLs for EPR pairs whose SL lies within 0.1~dB to 0.5~dB of the maximum SL reported for each platform.}
\label{figure6}
\end{figure}

\textit{Discussion.—}
We have established, and experimentally verified, the crucial role of both the mode structure and pump conditions in realizing spectrally uniform CV quantum microcombs. In the present implementation, we achieve raw two-mode squeezing up to $4.3~\mathrm{dB}$ and identify 14 EPR pairs with nearly uniform performance across a $0.7~\mathrm{THz}$ span. In Fig. 6, we compare the two-mode squeezing levels and the number of EPR pairs with other on-chip CV platforms. The CV quantum microcomb reported here simultaneously exhibits the highest on-chip vacuum two-mode squeezing level and the largest usable squeezed bandwidth to date.

Further dispersion engineering should increase the number of usable modes, while more efficient EO-comb sources will provide a larger set of LOs for homodyne detection \cite{yu2022integrated,lei2025strong}. Higher squeezing levels are likewise anticipated with improved overall efficiency enabled by chip-integrated filters and photodetectors \cite{stokowski2023integrated,larsen2025integrated,liu2025wafer}. The requisite few-transverse-mode structure, achieved here via a lossy metal ring that selectively damps higher-order families, could also be realized in intrinsically single-mode, high-$Q$ resonators as microfabrication technology advances. Together, these developments will pave the way for scalable sources of large-scale CV cluster states with highly uniform graph structure \cite{wang2025large,jia2025continuous}.

\textit {Acknowledgment}---This work was supported by Quantum Science and Technology-National Science and Technology Major Project (Grants No. 2024ZD0300800, 2023ZD0301100 and 2021ZD0301500), National Natural Science Foundation of China (12504430 and 62222515), and the High-performance Computing Platform of Peking University. Device fabrication in this work is supported by the Peking University Nano-Optoelectronic Fabrication Center and Synergetic Extreme Condition User Facility (SECUF).

\bibliography{main.bib}

\end{document}


\title{Supplemental Material to Perfect continuous-variable quantum microcombs}
\author
{Kangkang Li$^{1,\ast,\dagger}$, Yue Wang$^{1,\ast}$, Ze Wang$^{1,\ast}$, Xin Zhou$^{2,\ast}$, Jincheng Li$^{2,3}$, Yinke Cheng$^{1}$, Binyan Wu$^{1}$, Qihuang Gong$^{1,4,5,6}$, Bei-Bei Li$^{2,\dagger}$, Qi-Fan Yang$^{1,4,5,6,\dagger}$
\\$^{1}$State Key Laboratory for Artificial Microstructure and Mesoscopic Physics and Frontiers Science Center for Nano-optoelectronics School of Physics, Peking University, Beijing, 100871, China\\
$^2$Beijing National Laboratory for Condensed Matter Physics, Institute of Physics, Chinese Academy of Sciences, Beijing, 100190, China\\
$^3$School of Space and Earth Sciences, Beihang University, Beijing 100191, China\\
$^4$Collaborative Innovation Center of Extreme Optics, Shanxi University, Taiyuan, 030006, China\\
$^5$Peking University Yangtze Delta Institute of Optoelectronics, Nantong, Jiangsu, 226010, China\\
$^6$Hefei National Laboratory, Hefei, 230088, China\\
$^{\ast}$These authors contributed equally to this work.\\
$^{\dagger}$Corresponding author: kangkangli@pku.edu.cn; libeibei@iphy.ac.cn; leonardoyoung@pku.edu.cn.
}

\maketitle

\section{Theory} \label{sec1}
\subsection{Couple-mode equations of avoided mode crossings} \label{sec1}
For a Kerr microresonator supporting multiple transverse mode families, avoided mode crossings (AMX) are generally inevitable. These AMXs arise from coupling between distinct mode families, which perturbs the original modes and gives rise to characteristic spectral anti-crossings.
In our model, we consider two transverse mode families: the target mode family (denoted as $t$) with resonant frequencies given by $\omega^t_k=\omega_o^t+D_1^tk+\frac{D_2^t}{2}k^2+\mathcal{O}(k^3)$, and a crossing mode family (denoted as $c$) with resonant frequencies $\omega^c_k=\omega_o^c+D_1^ck+\mathcal{O}(k^2)$. Here, $D_1$ represents the free spectral range (FSR), $D_2$ is the second-order integrated dispersion and $k$ is the shared mode index.
The intersection of these two dispersion profiles is defined to occur at $k=k_0$.  

To characterize the quantum fluctuation of these two mode families, we introduce the annihilation operators $a_k$ for the target modes and $b_k$ for the crossing modes.
When the target mode family is pumped at $k=0$ under oscillation, the coupled equations of motion for $a_k$ and $b_k$ are given by
\begin{align}
    &\frac{d}{dt}a_k=-\left[\frac{\kappa^t}{2}+i(\delta\omega^t_k-2g|A_0|^2)\right]a_k+igA_0^2a_{-k}^{\dagger}+iGb_k+\sqrt{\kappa^t}V^t_{k}, \label{a}  \\
    &\frac{d}{dt}b_k=-\left(\frac{\kappa_k^c}{2}+i\delta\omega^c_k\right)b_k+iGa_k+\sqrt{\kappa^c}V^c_{k}  \label{b}  
\end{align}
where $G$ denotes the coupling strength between two mode families, $g$ is the nonlinear coupling coefficient, and $A_0$ represents the intracavity optical field amplitude.
$\delta\omega^t_k=\delta\omega_0+\frac{D^t_2}{2}k^2$ and $\delta\omega^c_k=\delta\omega_0+(k-k_0)(D_1^c-D^t_1)$ are the frequency detunings between $k_{\mathrm{th}}$ mode and the nearest resonant mode within the microresonator for the targeted and crossing mode respectively, where $\delta\omega_0$ is the pump-cavity detuning.
$\kappa^t$ and $\kappa^c$ separately denote the total loss rates for the target and crossing modes. 
The vacuum fluctuation associated with these channels are represented by the Langevin operators $\hat{V}^{t}_{k}$ and $\hat{V}^{c}_{k}$, which satisfy the following normalization relations
\begin{equation}
\begin{aligned}
    &\langle V_{k}^i(t)(V_{k'}^{i'}(t'))^{\dagger}\rangle=\delta_{i,i'}\delta_{k,k'}\delta(t-t'), \\
    &\langle V_{k}^i(t)V_{k'}^{i'}(t')\rangle=\langle (V_{k}^i(t))^{\dagger}V_{k'}^{i'}(t')\rangle=\langle (V_{k}^i(t))^{\dagger}(V_{k'}^{i'}(t'))^{\dagger}\rangle =0, \quad i,i'=t,c.
    \label{V}
\end{aligned}   
\end{equation}
Under the steady-state condition of the crossing modes, $\frac{d}{dt}b_k=0$, the relation between $b_k$ and $a_k$ is obtained as
\begin{equation}
   b_k=\frac{1}{\kappa^c/2+i\delta\omega^c_k}(iGa_k+\sqrt{\kappa^c}V^c_{k}) 
\end{equation}
Substituting this relation into Eq. (S1) eliminates $b_k$ and yields the following motion equation for $a_k$
\begin{equation}  
\frac{d}{dt}a_k=-\left[\frac{\kappa_k}{2}+i(\delta\omega_k-2g|A_0|^2)\right]a_k+igA_0^2a_{-k}^{\dagger}+\sqrt{\kappa_{k}}V_{k},
\end{equation}
where the effective loss rate $\kappa_k$, effective detuning $\delta\omega_k$ and combined vacuum noise operator $V_k$ are defined as
\begin{equation}
\kappa_k=\kappa^t+\frac{G^2\kappa^c}{(\kappa^c/2)^2+(\delta\omega^{c}_k)^2}, \;
\delta\omega_k=\delta\omega^t_k+\frac{G^2(-\delta\omega^{c}_k)}{(\kappa^c/2)^2+(\delta\omega^{c}_k)^2} , \;
V_{k}=\frac{1}{\sqrt{\kappa_{k}}}\left(\sqrt{\kappa^t}V^t_{k}+\frac{iG}{\kappa^c/2+i\delta\omega^c_k}\sqrt{\kappa^c}V^c_{k}\right).
\end{equation}
The redefined vacuum operator $V_k$ satisfies the normalization relations given in Eq. (S3). 
The perturbation induced by AMX on the loss and frequency are quantified by
\begin{equation}
     \Delta\kappa_{k}=\beta_k\kappa^c,\;
      \Delta\omega_k=\beta_k(-\delta\omega^{c}_k), 
  \quad  \mathrm{with} \;  \beta_k=\frac{G^2}{(\kappa^c/2)^2+(\delta\omega^{c}_k)^2}. 
    \label{mode shift}
\end{equation}
These results shows that when the total loss rate of the crossing modes is sufficiently large, satisfying $\kappa^c\gg 2G$, the AMX-induced perturbations are suppressed, i.e. $\Delta\omega_k\rightarrow0$ and $\Delta\kappa_k\rightarrow0$.
The following analysis focuses specifically on the frequency perturbations induced by AMXs, as this effect dominates in the situation considered in the main text.

\subsection{Two-mode squeezing with and without AMXs}
For the target mode family, the equations of motion for the annihilation operator pair $a_{\pm k}$ are given by  
\begin{equation}
\begin{aligned}
      \frac{d}{dt}\begin{pmatrix}
     a_{k}\\a^{\dagger}_{-k}
    \end{pmatrix}=
    J \begin{pmatrix}
      a_{k}\\a^{\dagger}_{-k}   
    \end{pmatrix}+\kappa \begin{pmatrix}
      V_{k}\\V^{\dagger}_{-k}   
    \end{pmatrix} , \quad \mathrm{with} \; 
    J=\begin{pmatrix}
     R_k & S_k \\ S_{-k}^{*} & R_{-k}^{*}
    \end{pmatrix} 
\end{aligned}
\label{var}
\end{equation}
where
\begin{equation}
\begin{aligned}
    R_{\pm k}=-\frac{\kappa}{2}-i(\delta\omega_{\pm k}-2g|A_0|^2), \;
    S_{\pm k}\equiv S= igA_0^2.
    \end{aligned}
\end{equation}
Here, $\kappa$ represents the total loss rate for $\pm k_{\mathrm{th}}$ qumodes and $\delta\omega_{\pm k}=\delta\omega_0+\frac{D_2}{2}k^2+\Delta\omega_{\pm k}$ denote their frequency detunings. The detunings include the pump-cavity detuning $\delta\omega_0$, second-order integrated dispersion $D_2$ and the frequency shift $\Delta\omega_{\pm k}$ induced by AMX.

For simplicity, we normalize the frequency quantities by $\kappa/2$ where $\kappa$ is the total loss rate at $k=0$, and taken $A_0$ to be real. The normalized form of Eq. (S8) is expressed as
\begin{equation}
    \frac{d}{d\tau}\begin{pmatrix}
     a_{k}\\a^{\dagger}_{-k}
    \end{pmatrix}= \begin{pmatrix}
    -1-i(\overline\zeta_k+\Delta_k-2\alpha) & i\alpha \\ -i\alpha & -1-+i(\overline\zeta_k-\Delta_k-2\alpha)
    \end{pmatrix}
    \begin{pmatrix}
      a_{k}\\a^{\dagger}_{-k}   
    \end{pmatrix}+ \begin{pmatrix}
    W_k \\ W_{-k}^{\dagger}
    \end{pmatrix}
\end{equation}
with the normalized quantities defined as
\begin{equation}
\begin{split}
    \tau=\frac{t}{\kappa/2},\quad \alpha=\frac{gA_0^2}{\kappa/2},\quad \zeta_{\pm k}=\frac{\delta\omega_{\pm k}}{\kappa/2} ,\quad W_{\pm k} =\sqrt{\frac{1}{\kappa_{\pm k}/2}} V_{\pm k}. 
\end{split}
\end{equation}
The average and asymmetric detunings are given by
\begin{equation}
    \overline \zeta_k=\frac{\zeta_k+\zeta_{-k}}{2}, \quad \Delta_k=\frac{\zeta_k-\zeta_{-k}}{2}.
\end{equation}

Considering the external coupling loss $\kappa_{e}$ (which is included in the total loss $\kappa$) and the detection efficiency $\eta_d$, along with their corresponding vacuum operators $V_{e,\pm k}$ and $V_{d, \pm k}$, the annihilation operators of the detected fields $c_k$ is derived as
\begin{equation}
    c_{\pm k}=\sqrt{\eta_d}(\sqrt{\kappa_{e}}a_{\pm k}-V_{e,\pm k})+\sqrt{1-\eta_d}V_{d,\pm k}
\end{equation}
Through Fourier transformation, at offset frequency $\omega \approx 0$, the shot-noise-normalized variance for the sum quadrature $x_k+x_{-k}$ ($x_k=c_k+c_k^{\dagger}$) is derived as 
\begin{equation}\label{h1}
\begin{aligned}
    \mathrm{Var}(x_k+x_{-k})=1-\eta \frac{2}{1+\sqrt{1+\frac{\Lambda_k}{4\alpha^2}}},   
\end{aligned}    
\end{equation}
with
\begin{equation}
     \eta=\eta_e\eta_d, \quad
     \Lambda_k=\left[1-\alpha^2+(\overline \zeta_k-2\alpha)^2-\Delta_k^2\right]^2+4\Delta_k^2.
\end{equation}
$\eta$ is the total efficiency, including extraction efficiency $\eta_e=\frac{\kappa_e}{\kappa}$ and detection efficiency $\eta_d$. The two-mode squeezing level (SL) for the pair $(k,-k)$ is given by $10\log_{10}[\mathrm{Var}(x_k+x_{-k})]$ in decibels (dB). 

In the presence of AMX at $k_{\mathrm{th}}$ qumode, the asymmetry in detuning becomes nonzero, i.e., $\Delta_k\neq 0$, as discussed in Section. IA.  
For small detuning satisfying $(\overline \zeta_k-2\alpha)^2<1+\alpha^2$, we have $\Lambda_k>\Delta_k^4+4$. The result indicates that $\Lambda_k$ increases rapidly with nonzero asymmetric detuning $\Delta_k$, leading to significant suppression of squeezing, while the minimal value of $\Lambda_k$, corresponding to the highest squeezing, occurs at symmetric detuning $\Delta_k=0$. 

In the absence of AMX, the asymmetries disappear, i.e. $\Delta_k=0$, along with $\overline\zeta_k=\zeta_k$. Therefore, the quadrature-noise variance without AMX is simplified as
\begin{equation}
\begin{aligned}
    \mathrm{Var}(x_k+x_{-k})=1-\eta\frac{2}{1+\sqrt{1+\frac{\Lambda_k}{4\alpha^2}}}, 
    \quad  \mathrm{with} \; 
    \Lambda_k=\bigr[1-\alpha^2+(\zeta_k-2\alpha)^2\bigr]^2 
    \label{tms}
\end{aligned}
\end{equation}
The minimal variance is achieved at $\zeta_k-2\alpha=0$.

\section{Device}
\subsection{Device design}
As discussed in Section. I A, the suppression of the perturbations induced by AMXs requires a sufficiently high loss rate of the crossing mode family. To achieve this, we employed a modified metal ring structure on the silica microdisk to implement dissipation engineering, which selectively dampens higher-order modes. Figure S1(a) presents the simulation results of the highest intrinsic quality factor $Q_i$ and the number of modes per FSR with $Q_i>5\times 10^6$ as a function of the ring displacement $d$. The results indicates that a displacement of $d\approx 15$ $\mu$m yields both the highest $Q_i$ and and a relatively small number of high-$Q_i$ modes, making it the optimal configuration.

Figure S1(b) shows the stimulated integrated dispersion of the disk both without and with the metal ring at the optimal distance. 
By setting the external coupling quality $Q_e$ of different mode families to match the intrinsic quality of the fundamental mode at 1550 nm ($Q_i^0$), the normalized transmission reveals the $Q_i$ values of individual modes, where 0 and 1 corresponds to $Q_i=Q_i^0$ and $Q_i\ll Q_i^0$ respectively. 
The results demonstrate that the metal-modified microdisk significantly suppresses AMXs, particularly for the fundamental mode family, owing to the strong dissipation of higher-order modes.

\begin{figure*}[h]
\centering
\includegraphics[width=1.0\linewidth]{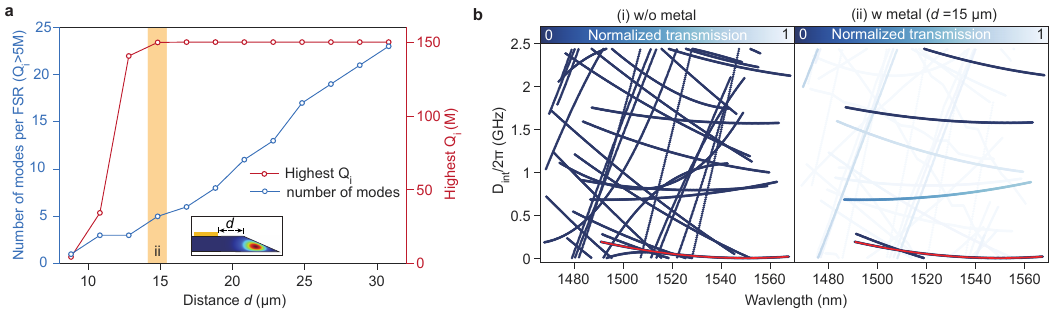}
\caption{{\bf Simulated characteristics of the metal-modified microdisk.} {\bf a,} Stimulated highest $Q_i$ (blue) and number of modes per FSR with $Q_i>5\times 10^6$ (red) versus the ring displacement $d$. Orange shading corresponds to the optimal $d$ (case b(ii)) with maximum highest $Q_i$ and relatively small number of high-$Q_i$ modes. M refers to million. {\bf b,} Stimulated integrated dispersion of different mode families in the absence of mental (i) and in the presence of mental with optimal $d$ (ii). The red curve marks the fundamental mode family. $Q_e$ of all modes is set same as the $Q_i$ of the fundamental mode at 1550 nm ($Q_{i}^0$).}
\label{figureS1}
\end{figure*}

\subsection{Device fabrication}
Based on a 6 $\mu$m-thick thermal $\mathrm{SiO_2}$ layer on a silicon substrate, silica microdisks featuring a wedged edge of approximately $30^{\circ}$ were fabricated using photolithography and hydrofluoric (HF) wet etching. A second photolithography step was then employed to transfer a concentric metal ring pattern onto the microdisks. A 150 nm-thick aluminum (Al) layer was first deposited via electron beam physical vapor deposition, followed by a 20 nm-thick gold (Au) layer. After removing the excess metal through the lift-off process, suspended microdisks with an undercut region of about 100 $\mu$m were produced through isotropic etching of the silicon pedestal using xenon difluoride (XeF$_2$). Here, the Al layer exhibits strong adhesion to silica and a high absorption coefficient of $1.145\times 10^6 \mathrm{cm}^{-1}$, enabling effective suppression of AMX induced by higher-order transverse modes. The Au layer serves as a protective coating to prevent oxidation of the underlying Al and maintains surface hydrophobicity, thereby reducing surface roughness during subsequent etching processes.

\subsection{Characterization of the extraction efficiencies for the transverse mode family}
The extraction efficiencies of the transverse mode family is characterized using a tunable continuous-wave laser (New Focus, TLB-6700). The relative frequency of the laser is calibrated using a Mach-Zehnder interferometer.
As shown in Fig. S2, the extraction efficiencies remain consistently high at approximately 90\% across the entire spectral range utilized in this study, spanning modes -50 to 50 centered at 1543.2 nm.

\begin{figure*}[hbtp]
\centering
\includegraphics[width=0.5\linewidth]{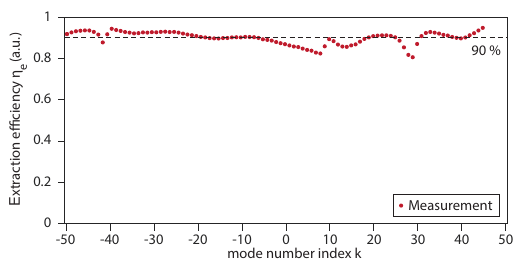}
\caption{{\bf Extraction efficiencies distribution.} 
Distribution of extraction efficiency $\eta_e$ of the transverse mode family against the mode index. The $0_{\mathrm{th}}$ mode is located at 1543.2 nm.}
\label{figureS2}
\end{figure*}

\section{Experimental details}
\subsection{Detailed experimental setup}
The experimental setup is presented in Fig.3. 
The electro-optic comb (EO comb) is generated via phase and amplitude modulations of a continuous-wave laser (New Focus, TLB-6700). The phase and intensity modulators are driven by the amplified low-noise 25 GHz microwave signals from a microwave generator (Keysight PSG E8257D), with their phases optimized to produce a braod and flat spectrum. A small portion (10\%) of the EO comb is tapped and monitored by a photodetector to servo-lock the direct current (DC) bias of the intensity modulator for ensuring long-term stability. A multi-channel programmable filter (II-IV Waveshaper 4000A) with a spectral resolution of 10 GHz is employed to synthesize the pump and local oscillator (LO) from the EO comb. By applying appropriate group velocity dispersion through the programmable filter, the temporal width of the comb is minimized, resulting in a flat phase profile across all comb lines. 

The optical signal transmitted from the microresonator and LO light are interfered on a 50:50 fiber directional coupler, achieving an interference efficiency of 99\%. Subsequently, two band-pass filters with a bandwidth of 500 GHz and a extinction ratio of 20 dB are used to remove pump lines, while maintaining a transmission efficiency of approximately 95\% for other wavelengths. The filtered signal is then detected by a high-quantum-efficiency (99\%) balanced homodyne photodetector (Laser Components, IGHQEX0500) with an input power of 2.2 mW and a common-mode noise rejection ratio of 35 dB. The quadrature-noise spectra are then recorded using an electrical spectrum analyzer (ESA, Rohde \& Schwarz FPL1026). The overall detection efficiency of the system is calculated to be 78\%, taking into account an optical loss of 17\%, the photodiode quantum efficiency of 99\%, and the transmission of the tapered fiber of 95\%. 

\subsection{Calibration of the pump power and pump detuning}
To compare the bandwidth of continous-variable (CV) quantum microcombs at identical intracavity power (corresponding to $\alpha=0.8$ as requirement) for different normalized pump detunings ($\zeta_0=0.4$ and $\zeta_0=0$), the pump conditions were precisely calibrated prior to experimental characterization. Due to the upper limit of pump power available in our system, we slightly reduced the loaded quality to $Q\approx1.6\times 10^7$ for lower oscillation threshold, with the corresponding extraction efficiency reduced to $\eta_e\approx0.85$.

We first determined the required pump powers using the established relation for the normalized pump power $f^2$ under oscillation\cite{godey2014stability}
\begin{equation}
    f^2=\alpha[1+(\zeta_0-\alpha)^2]
\end{equation}
For $\alpha=0.8$ with $\zeta_0=0.4$ and $\zeta_0=0$, the corresponding normalized pump powers are determined as $f^2=0.9$ and $f^2=1.3$ respectively. The oscillation threshold, identified as the point where the comb power is fully eliminated, is measured to be $P_{\mathrm{th}}=32~\mu$W. Thus, the corresponding pump powers, calculated as $f^2P_{\mathrm{th}}$, are 29~$\mu$W and 39~$\mu$W, respectively.

To maintain $\alpha=0.8$, we injected calibrated pump powers into the Kerr microresonator and adjusted the pump detunings by monitoring the normalized transmission $T$. The combination of input-output relation $A_{\mathrm{out}}=A_{\mathrm{in}}-\sqrt{\kappa_e}A_0$ and input-intracavity relation $A_{0}=\frac{\sqrt{\kappa_e}}{\kappa/2+i(\delta\omega_0-|A_0|^2)}A_{\mathrm{in}}$ in the Kerr microresonator under oscillation, yields the expression of normalized transmission
\begin{equation}
    T=\frac{|A_{\mathrm{out}}|^2}{|A_{\mathrm{in}}|^2}=1-\frac{4\eta_e(1-\eta_e)}{1+(\zeta_0-\alpha)^2}
\end{equation}
Therefore, the normalized transmissions at $\alpha=0.8$ for $\zeta_0=0.4$ and $\zeta_0=0$ are $T=0.57$ and $T=0.69$ respectively. By stabilizing the system at the corresponding operating points shown in Fig.5, we successfully established the required pumping conditions.

\bibliography{SM.bib}